\documentclass[a4paper,11pt]{article}
\pdfoutput=1 % if your are submitting a pdflatex (i.e. if you have
\usepackage{jheppub}

\usepackage[T1]{fontenc} % if needed

\usepackage{amsfonts,amssymb,amsmath,graphicx}
\usepackage{color}

\preprint{DFPD/2017/TH/12}

\title{\boldmath Extremal Black Holes, Stueckelberg Scalars and Phase Transitions}

\author[a,b]{Alessio Marrani,}
\author[c]{Olivera Miskovic,}
\author[c]{and Paula Quezada Leon}

\affiliation[a]{Museo Storico della Fisica e Centro Studi e Ricerche ``Enrico Fermi'',\\Via Panisperna 89A, I-00184, Roma, Italy}
\affiliation[b]{Dipartimento di Fisica e Astronomia ``Galileo Galilei'', Universit\`{a} di Padova,\\
and INFN, Sez. di Padova,Via Marzolo 8, I-35131 Padova, Italy}
\affiliation[c]{Instituto de F\'{\i}sica, Pontificia Universidad Cat\'{o}lica de Valpara\'{\i}so,\\Casilla 4059, Valpara\'{\i}so, Chile}

\emailAdd{alessio.marrani@pd.infn.it}
\emailAdd{olivera.miskovic@pucv.cl}
\emailAdd{pquezada.l@gmail.com}

\abstract{We calculate the entropy of a static extremal black hole in 4D gravity, non-linearly
coupled to a massive Stueckelberg scalar. We find that the scalar field does not allow the black hole to be magnetically charged.
We also show that the system can exhibit a phase transition due to electric charge variations. For spherical and hyperbolic horizons, the critical point exists only in presence of a cosmological constant, and if the scalar is massive and non-linearly coupled to electromagnetic field. On one side of the critical point, two extremal solutions coexist: Reissner-Nordstr\"{o}m (A)dS black hole and
the charged hairy (A)dS black hole, while on the other side of the critical point the black hole does not have hair. A near-critical analysis reveals that the hairy black hole has larger entropy, thus giving rise to a zero temperature phase transition. This is characterized by a discontinuous second derivative of the entropy with respect to the electric charge at the critical point. The results obtained here are analytical and based on the entropy function formalism and the second law of thermodynamics.}

\begin{document}
\maketitle
\flushbottom

%%%%%%%%%%%%%%%%%%%%%%%%%%%%%%%%%%%%%%%%%%%%%%%%%%%%%%%%%%%%%%%%%%%%%%%%%%%%%%%%%%%%%%%%%%%%%%%%%%%%%%%%%%

\section{Introduction}

The AdS/CFT correspondence has brought a novel approach to field theories, allowing to obtain results in the strong coupling regime from
gravitational computations in a classical approximation. A remarkable cornerstone concerns the description of phase transitions in a thermal field theory beyond the applicability of the BCS model. A holographic dual of a thermodynamical system at equilibrium, at constant temperature $T$,
has a minimal internal energy and, in the AdS gravity side, it corresponds to a black hole configuration with Hawking temperature $T$.
For instance, a holographic superconductor in four dimensions is dual to a five-dimensional charged AdS black hole coupled to a complex scalar field \cite%
{Hartnoll:2008vx,Hartnoll:2008kx,Gubser:2008px}. A phase transition in a QFT$_{4}$ occurs when a charged black hole develops a scalar hair below some
critical temperature. This model successfully explains the large energy gap of high-temperature superconductors, with gravitational computations correctly yielding the transport coefficients of the superconductor.

However, the aforementioned method, based on a calculation of the free energy from the Euclidean action, fails at zero temperature or, on the gravity side, when the black hole becomes extremal. It is worth here remarking that a zero temperature system can also be realized as a soliton configuration, and a phase transition between an AdS black hole and an AdS soliton has been discussed in Refs. \cite{Nishioka et al,Horowitz et al} at any temperature, including $T=0$ as a limit case; thence, the dual QFT describes a transition superconductor/insulator.

In this work, we focus on extremal black holes. An analogue of the Meissner
effect, typical of phase transitions, was observed in extremal Kerr and Kerr-Newman black holes embedded in an external
magnetic field  \cite{Astorino2015}. Near the horizon, the magnetic field can be reabsorbed in
a redefinition of the parameters, so the magnetic field can be considered
as expelled from the near-horizon geometry, when the Hawking temperature is
zero. A similar treatment was given in \cite{Bicak} and, including acceleration, in \cite{Astorino2016}.

Since extremal black holes have zero temperature and they are not described by conventional thermodynamics, an equilibrium state of a thermodynamical system is more suitably described as a maximum entropy state in the entropy representation of states; subsequently, the entropy of the extremal black hole arises as a consequence of the degeneracy of the quantum ground state. This phenomenon is known in the physics of condensed matter (spin glasses).

The macroscopic entropy of an extremal black hole can be calculated using the entropy function formalism \cite{Sen:2005wa, Sen:2005iz}. In $D$ dimensions, this is based on a variational principle applied to a generic class of entropy functions of the
charges, scalar fields and parameters of the near-horizon geometry  AdS$_{2}\times \mathbb{S}^{D-2}$,
AdS$_{2}\times \mathbb{R}^{D-2}$, AdS$_{2}\times \mathbb{H}^{D-2}$ (in static cases) or AdS$_{2}\times U(1)$ (rotating case).
Remarkably, the formalism can be generalized to other
extremal geometries, such as warped ones \cite{Astefanesei:2012ar}.
The extremization of the entropy function determines all the near-horizon
parameters, thus enabling to extract information about the black holes without knowledge of a particular solution.
In addition, the AdS/CFT correspondence also suggests that the horizon geometry contains all the data available in the bulk of the spacetime, including its asymptotic region. Indeed, the entropy function can be derived from purely boundary term of the gravitational action \cite{BibhasRanjanMajhi}.

On the other hand, phase transitions due to thermal or, as in our case,
electric charge fluctuations, are related to instabilities
of the system around a critical point. In particular, it has been noticed
that a massless scalar field produces an instability at the horizon of an
extreme Reissner-Nordstr\"{o}m (RN) black hole \cite{Aretakis2013} and that the
axisymmetric extremal horizons are unstable under linear scalar
perturbations \cite{Aretakis2015}. Similar instabilities also occur for a
massive scalar field \cite{Lucietti-Murata-Reall}. Recently, these
instabilities were studied in the extremal case, by performing an analysis of the charged scalar perturbations for RN and Kerr
solutions \cite{Zimmerman:2016qtn}.

Zero temperature phase transitions of a four-dimensional charged black hole would occur in General Relativity, as well, when it is
non-minimally coupled to a Stueckelberg scalar \cite{Stuckelberg}. In fact, non-linear Stueckelberg interactions have been known to describe
both first and second order thermal phase transitions \cite{Hartnoll:2008kx,Franco:2009yz}.
A question we address here is whether a similar transition would also occur at
zero temperature.

\medskip

The plan of the paper is as follows.

In Sec. 2 we introduce the entropy function formalism, then apply it to the Stueckelberg system in Sec. 3. The case of vanishing scalar at the horizon (\textit{e.g.}, no scalar hair) is treated in Sec. 4, whereas the hairy black hole is investigated in Sec. 5. A detailed analysis in vicinity of the critical point is provided in Sec. 6, and it yields to the first evidence of a phase transition for extremal black holes. Final comments and an outlook are given in Sec. 7.

\section{Entropy function of the black hole}

The extremal black hole is the smallest mass black hole for a given electric
charge and angular momentum in flat space, but in curved
space (such as AdS space) it can be modified due to the renormalization of the stress tensor \cite{Balasubramanian-Kraus,Miskovic-Olea}.
Geometrically, it has two horizons that overlap, what implies that its
near-horizon geometry in four space-time dimensions has the topology
$\mathbb{H}\simeq \,$AdS$_{2}\times \Sigma _{k}$,
where AdS$_{2}$ is two-dimensional anti-de Sitter space and $\Sigma _{k}$
is a 2-dimensional constant curvature space of the unit radius $k$.
For $\Lambda =0$ and $\Lambda >0$, the horizon is always spherical ($k=1$) and we
have that $\Sigma _1\simeq \mathbb{S}^2$ is a two-sphere, whereas for
$\Lambda <0$ there are three possibilities, the spherical horizon black hole
with $\Sigma _1\simeq \mathbb{S}^{2}$, planar black hole with
$\Sigma_0\simeq \mathbb{R}^{2}$ and a hyperbolic horizon black hole with
$\Sigma _{-1}\simeq \mathbb{H}^{2}$. Let $v_{1}$ and $v_{2}$ be the radii of
the AdS$_{2}$ and $\Sigma _{k}$ sections, respectively. Then the horizon of
the black hole in spherical coordinates is described by the metric
\begin{equation}
\left. ds^{2}\right\vert _{\mathbb{H}}=g_{\mu \nu }(x)\,dx^{\mu }dx^{\nu
}=v_{1}\left( -r^{2}\,dt^{2}+\frac{dr^{2}}{r^{2}}\right) +v_{2}\,%
d\Omega_{(k)}^2\,,  \label{metric}
\end{equation}
where $r$ is the radial distance from the horizon
and the transversal section
$\Sigma _k$ is given by the square interval
\begin{equation}
d\Omega_{(k)}^2=\gamma_{mn}(y)\,dy^{m}y^{n}\,,
\end{equation}
where $y^m$ ($m=3,4$) are the local coordinates on $\Sigma_k$. Explicitly,
\begin{equation}
\begin{array}{lll}
d\Omega _{(1)}^{2} & =d\theta ^{2}+\sin^2 \theta \,d\varphi ^{2}\,,\qquad  &
y^{m}=(\theta ,\varphi )\,, \\
d\Omega _{(0)}^{2} & =dx^{2}+dy^{2}\,, & y^{m}=(x,y)\,, \\
d\Omega _{(-1)}^{2} & =d\chi ^{2}+\sinh^2 \chi \,d\lambda ^{2}\,, & y^{m}=(\chi,\lambda )\,.
\end{array}
\label{transversal}
\end{equation}
The invariant volume element of $\Sigma_k$ is given by $\sqrt{\gamma}\,d^{2}y$,
with $\sqrt{\gamma }=\sin \theta $, $1$, $\sinh \chi $, for $k=1$, $0$, $-1$, respectively.

The metric (\ref{metric}) corresponds to a near-horizon limit of some
\textit{static} black hole coupled to an electromagnetic field $A_{\mu }(x)$
and to a scalar field $\phi (x)$, described by the action
\begin{equation}
I=\int d^{4}x\sqrt{-g}\,\mathcal{L}(g,A,\phi )\,.  \label{action}
\end{equation}
Near the horizon, the metric behaves as (\ref{metric}),
and the electromagnetic field, described
by the field strength $F_{\mu \nu }=\partial _{\mu }A_{\nu }-\partial _{\nu }A_{\mu }$,
is characterized by the electric charge $e$ and the magnetic charge $p$, such that
\begin{equation}
F_{rt}=e\,,\qquad F_{34}=\frac{p}{4\pi }\,\sqrt{\gamma }\,, \label{EM}
\end{equation}
in the coordinates (\ref{metric}). On the other hand, the scalar field, due to the
attractor mechanism, does not depend on its asymptotic value, but only on the
horizon value, $u$.  Thus, on the horizon, the fields are represented by a
set of parameters
\begin{equation}
\mathbb{H}:\qquad g_{\mu \nu }\rightarrow (v_{1}\,,\,v_{2})\,, \quad
A_{\mu}\rightarrow (e\,,p)\,,\quad \phi \rightarrow u\,.
\label{near H}
\end{equation}
The action (\ref{action}) evaluated on the horizon is given by an auxiliary
function
\begin{equation}
f(v_1,v_2,e,p,u)=\int\limits_{\mathbb{H%
}}d\theta d\phi \sqrt{-g}\,\mathcal{L}(v_1,v_2,e,p,u)\,,  \label{f}
\end{equation}
which satisfies the action principle, that is, has an extremum on the
equations of motion, for given boundary conditions. The entropy function
$\mathcal{E}(v_1,v_2,e,p,u)$ is the Legendre transformation of the function $f$ with respect to the electric
field,
\begin{equation}
\mathcal{E}(v_1,v_2,e,p,u)=2\pi [eq-f(v_1,v_2,e,p,u)] \,,
\label{entropy function}
\end{equation}
where $q$ is the asymptotic electric charge. The parameters near the horizon
are calculated as an extremum of the entropy function \cite%
{Sen:2005wa,Sen:2005iz},
\begin{equation}
\frac{\partial \mathcal{E}}{\partial v_{i}}=0\text{\quad }(i=1,2)\,,\qquad
\frac{\partial \mathcal{E}}{\partial u}=0\,,\qquad \frac{\partial \mathcal{E}%
}{\partial e}=0\,, \qquad \frac{\partial
\mathcal{E}}{\partial p}=0\,,   \label{extremum}
\end{equation}
and the black hole entropy $S$ is its extremal value,
\begin{equation}
S=\mathcal{E}_{\mathrm{extr}}\,.
\end{equation}

Therefore, finding the entropy function $\mathcal{E}(v_1,v_2,e,p,u)$ and its maximum, one
can calculate the entropy, electric field, AdS$_{2}$ and $\Sigma _{k}$ radii
 of the extremal black hole, independently on a particular static solution under
consideration.

For $k=1$, $v_{1}=v_{2}$ yields the famous Bertotti-Robinson near-horizon
conformally flat metric \cite{BR1,BR2}, which characterizes extremal black holes
in the ungauged theory (also, for the interesting case of Abelian ``flat gaugings'', see \cite{Katm}).

\section{Black hole coupled to a Stueckelberg scalar}

Consider General Relativity with a cosmological constant $\Lambda $ coupled
to electromagnetic and scalar fields,\footnote{Here $g_{\mu \nu }$ denotes the full spacetime metric in
arbitrary coordinates and it will be chosen in the particular form (\ref{metric}) when evaluated on the horizon.
These two metrics cannot be confused.}
\begin{equation}
I=\int d^{4}x\sqrt{-g}\,\left\{ \frac{1}{16\pi G_N}\,(R-2\Lambda ) -\frac{1}{%
4}\,F_{\mu \nu }F^{\mu \nu }-\frac{1}{2}\, \left[ (\partial
\phi)^{2}+m^{2}\phi ^{2}+\mathcal{F}(\phi )(\partial \sigma -A)^{2}\right]
\right\} \,,  \label{I}
\end{equation}
where $R=g^{\mu \nu }R_{\ \mu \alpha \nu }^{\alpha }$ is the scalar
curvature. Stueckelberg complex scalar $\hat{\phi}(x)=\phi (x)\,$e$%
^{i\sigma (x)}$ (with $\phi$ and $\sigma$ real) is minimally coupled when
the Stueckelberg function is quadratic, $\mathcal{F}(\phi )=\phi ^{2}$.
Non-minimal interaction has to satisfy $\mathcal{F}(\phi )>0$, $\phi \neq 0$%
, and $\mathcal{F}(0)=0$ \cite{Stuckelberg}. Such couplings preserve an $U(1)
$ invariance and lead to first and second order phase transitions in
non-extremal cases \cite{Hartnoll:2008kx,Franco:2009yz}. In the extremal
case, non-linear terms in $\mathcal{F}$ might produce instabilities at the
horizon \cite{Aretakis2013}, so we choose
\begin{equation}
\mathcal{F}(\phi )=\phi ^{2}+\frac{a}{4}\,\phi ^{4}\geq 0\,,
\label{Stuckelberg}
\end{equation}
where $a$ is some coupling constant of dimension (length)$^{2}$ in natural
units. Note that, in general, the constant $a$ can be positive or negative.

Equations of motion obtained from the action (\ref{I}) read
\begin{eqnarray}
\delta g^{\mu \nu } &:& \qquad R_{\mu \nu }-\frac{1}{2}\,g_{\mu \nu}\,R-g_{\mu \nu }\,\Lambda
=8\pi G_{N}\,T_{\mu \nu }\,,  \nonumber \\
\delta A_{\mu } &:&\qquad \nabla _{\mu }F^{\mu \nu }
=\mathcal{F}(\phi )\,(\nabla^{\mu }\sigma -A^{\mu }) \,,  \nonumber \\
\delta \phi  &:&\qquad (\square -m^{2}) \phi
=\frac{1}{2}\,\frac{d\mathcal{F}}{d\phi}\,(\nabla \sigma-A)^{2}\,,  \nonumber \\
\delta \sigma  &:&\qquad \nabla _{\mu }\left[ \mathcal{F}(\phi )\,
(\nabla ^{\mu}\sigma -A^{\mu }) \right] =0\,,
\end{eqnarray}
where $\nabla _{\mu }$ is a covariant derivative with respect to the affine
connection associated to the metric $g_{\mu \nu }$. The energy-momentum
tensor of the matter has the form
\begin{eqnarray}
T_{\mu \nu } &=&F_{\mu \lambda }F_{\nu }^{\ \lambda }-\frac{1}{4}\,g_{\mu\nu }
\,F^{2}+\partial _{\mu }\phi \partial _{\nu }\phi +\mathcal{F}(\phi )(\partial_{\mu }
\sigma -A_{\mu })(\partial _{\nu }\sigma -A_{\nu })  \nonumber \\
&&-\frac{1}{2}\,g_{\mu \nu }\left[ (\partial \phi )^{2}
+m^{2}\phi^{2}+\mathcal{F}(\phi )(\partial \sigma -A)^{2}\right] \,.
\end{eqnarray}
A non-negative energy density of the matter is ensured by imposing the
weak-energy condition $-T_{\mu \nu }u^{\mu }u^{\nu }\geq 0$, for a timelike
unit vector $u^{\mu }$.

The equation of motion for the field $\sigma (x)$ is not independent, but it
can be obtained from a divergence of the equation for $A_{\mu }$. This is a
consequence of $U(1)$ gauge symmetry because $\sigma (x)$ is a nonphysical
parameter, which can be consistently eliminated by the gauge-fixing $\sigma=0$.

On the horizon $\mathbb{H}$, spherically symmetric static field
configurations are replaced by
five parameters $(v_{1},v_{2},e,p,u)$, where
the gauge fixing $\sigma=0$ has been taken into account.
The electromagnetic field (\ref{EM}) can be obtained from the electromagnetic potential
with non-vanishing components $A_t=er$ and $A_4(x^3)$ which, when $k=1,0,-1$,
can be chosen as $A_{\varphi }=-\frac{p}{4\pi }\,\cos \theta $, $A_{y}=\frac{p}{4\pi }\,x$,
$A_{\lambda }=\frac{p}{4\pi }\,\cosh \chi $, respectively.

The parameters $(v_{1},v_{2},e,p,u)$ are taken to be independent, as they are coming from the independent fields in the action. A condition of the extremum of the entropy function will introduce relations among them, and they will become dependent. This method includes a broad class of asymptotic conditions on the fields, which do not introduce constraints on the near-horizon parameters.

The Lagrangian density (\ref{I}) evaluated on $\mathbb{H}$ reads
\begin{equation}
\mathcal{L}=\frac{1}{8\pi G_{N}}\left(\frac{k}{v_{2}}-%
\frac{1}{v_{1}}-\Lambda \right) -\frac{1}{2}\left(\frac{p^{2}}{16\pi ^{2}v_{2}^{2}}
-\frac{e^{2}}{v_{1}^{2}}\right) -\frac{1}{2%
}\,m^{2}u^{2}-\frac{1}{2}\,\mathcal{F}(u)\, \left(
\frac{p^{2}z_{k}}{16\pi ^{2}v_{2}}-\frac{e^{2}}{v_{1}}\right) \,,  \label{L}
\end{equation}
where for the scalar field we set $\phi=u$, $\partial _{\mu }\phi =0$.
The function $z_{k}(y^{m})$ depends on the horizon geometry and is
given by $z_{1}=\cot ^{2}\theta $, $z_{0}=x^{2}$ and $z_{-1}=\coth ^{2}\chi $.

The auxiliary function $f=\int d^{2}y\sqrt{\gamma }\,v_{1}v_{2}\mathcal{L}$ is
obtained after integrating out Eq.(\ref{L}) over the transversal section,
\begin{eqnarray}
f &=&\text{Vol}(\Sigma _{k})\,v_{1}v_{2}\left[ \frac{1}{8\pi G_{N}}
\left(\frac{k}{v_{2}}-\frac{1}{v_{1}}-\Lambda \right)
-\frac{1}{2}\left( \frac{p^{2}}{16\pi ^{2}v_{2}^{2}}-\frac{e^{2}}{v_{1}^{2}}\right)
-\frac{1}{2}\,\left( m^{2}u^{2}-\frac{e^{2}}{v_{1}}\,\mathcal{F}(u)\right) \right]   \nonumber \\
&&-\frac{1}{2}\,\mathcal{F}(u)\,\frac{v_{1}v_{2}p^{2}}{16\pi ^{2}v_{2}}\,
\int d^{2}y\sqrt{\gamma }\,z_{k}(y^{m})\,,
\end{eqnarray}
where we denoted the horizon volume by Vol$(\Sigma _{k})=\int d^{2}y\sqrt{\gamma }$.
When $k=1$, the horizon is compact and Vol$(\Sigma _{k})=4\pi $, and when $k=0$ or $-1$,
the horizon is non-compact and Vol$(\Sigma _{k})$ is infinite.
In those cases, the physical quantity of interest is the entropy density
(entropy per unit volume of the horizon).

The only explicit dependence on $y^{m}$ in $\mathcal{L}$ is through $z_{k}$, and
straightforward calculation shows that it is divergent. Thus, the only possibility to
have a finite solution is to set the magnetic charge to zero, $p=0$. We conclude that, when $\phi=0$, the RN black hole can be magnetically charged; however, when $\phi\neq 0$, the magnetic field $F_{mn}\neq 0$ breaks
the spherical symmetry of the solution. Thus, to allow for hairy black holes, we will henceforth set $p=0$.

With this at hand and with the vanishing magnetic charge,
the entropy function (\ref{entropy function}) reads
\begin{equation}
\mathcal{E}=2\pi eq-2\pi \text{Vol}(\Sigma _{k})\,v_{1}v_{2}%
\left[ \frac{1}{8\pi G_{N}} \left(\frac{k}{v_{2}}-\frac{1}{%
v_{1}}-\Lambda \right) +\frac{e^{2}}{2v_{1}^{2}}-\frac{1}{2}\,\left(
m^{2}u^{2}-\frac{e^{2}}{v_{1}}\,\mathcal{F}(u)\right)\right] \,.
\label{entropy f-on}
\end{equation}

An extremum of the above function gives rise to the following equations of
motion,
\begin{equation}
\begin{array}[b]{lll}
0=\dfrac{\partial \mathcal{E}}{\partial v_{1}} & \quad \Rightarrow \quad & %
k-\Lambda v_{2}=v_{2}\left( \dfrac{e^{2}}{v_{1}^{2}}%
+m^{2}u^{2}\right)\,,\medskip \\
0=\dfrac{\partial \mathcal{E}}{\partial v_{2}} & \quad \Rightarrow \quad &
1+\Lambda v_{1}=\dfrac{e^{2}}{v_{1}}-v_{1}\,m^{2}u^{2}+e^{2}\mathcal{F}%
(u)\,,\medskip \\
0=\dfrac{\partial \mathcal{E}}{\partial e} & \quad \Rightarrow \quad &
Q=v_{2}e\left( \dfrac{1}{v_{1}}+\mathcal{F}(u)\right) \,,\medskip \\
0=\dfrac{\partial \mathcal{E}}{\partial u} & \quad \Rightarrow \quad &
0=v_{2}\left( 2v_{1}\,m^{2}u-e^{2}\,\dfrac{d\mathcal{F}}{du}\right) \,,%
\end{array}
\label{eom}
\end{equation}
where we set $4\pi G_{N}=1$ for the sake of simplicity and also %
introduced the charge density $Q=\frac{q}{\text{Vol}(\Sigma
_{k})}$. Eqs.(\ref{eom}) are invariant under the reflection $(e,Q)
\rightarrow (-e,-Q)$, and $e$ and $Q$ have the same sign (see third
equation) so we can choose, without loss of generality, $e$, $Q>0$.

Equations (\ref{eom}) have different branches of solutions that have to be
discussed independently.

\section{Extremal black hole without hair \label{NonHairy}}

$u=0$ is always a particular solution of the scalar equation in (\ref{eom}),
thus we first focus to that case. We know that the result is the Reissner-Nordstr\"{o}m flat, dS or AdS black hole, but for consistency we employ the entropy function method.

We have to solve the system
\begin{eqnarray}
k-\Lambda v_{2}&=& \frac{e^{2}v_{2}}{v_{1}^{2}}\,, \qquad Q=\frac{ev_{2}}{v_{1}}\,, \nonumber\\
1+\Lambda v_{1}&=&\frac{e^{2}}{v_{1}}\,.
\label{eom u=0}
\end{eqnarray}

When $k=0$, a general solution of the above algebraic equations is
\begin{eqnarray}
v_{1}^{0} &=&-\frac{1}{2\Lambda }\,, \qquad v_{2}^{0} =\frac{Q}{\sqrt{-\Lambda }}\,, \notag \\
e^{0} &=&\frac{1}{2\sqrt{-\Lambda }}\,. \label{planar}
\end{eqnarray}
It exists only when $\Lambda <0$, as expected for the planar horizons. The entropy $S_{k=0}(Q)$ of this black hole is the entropy function evaluated on (\ref{planar}), obtaining in that way the entropy density as
\begin{equation}
s_{0}(Q)=\frac{S_{0}(Q)}{\text{Vol}(\Sigma _{0})}=\frac{\pi Q}{\sqrt{-\Lambda }}\,.
\end{equation}

When $k=\pm 1$, the general solution of
Eqs.(\ref{eom u=0}) for fixed $Q$ exists provided $\Delta =1-4\Lambda
Q^{2}>0$, and it reads
\begin{eqnarray}
v_{1}^{\pm }(Q) &=&\frac{2Q^{2}}{1-4\Lambda Q^{2}\pm \sqrt{1-4\Lambda Q^{2}}}%
\,,  \notag \\
v_{2}^{\pm }(Q) &=&\frac{2Q^{2}k}{1\pm \sqrt{1-4\Lambda
Q^{2}}}\,,  \notag \\
e^{\pm }(Q) &=&\frac{Q}{\sqrt{1-4\Lambda Q^{2}}}\,.  \label{solution u=0}
\end{eqnarray}
Other inequalities that have to be fulfilled are
\begin{equation}
1+\Lambda v_1>0\,,\qquad k(1+2\Lambda v_1)>0\,.
\label{2inequalities}
\end{equation}
Using the identity $\Lambda v_{1}^{\pm }
=\frac{1}{2}\left( -1\pm \frac{1}{\sqrt{\Delta }}\right) $, we find that the inequalities are satisfied
depending on the geometry of the horizon, so that
\[
\begin{array}{lll}
k=+1\,, & \Delta >0\,,\quad  & \text{solution is }(v_{1}^{+},v_{2}^{+},e^{+})\,, \\
k=-1\,,\qquad  & \Delta >1\,,\quad  & \text{solution is }(v_{1}^{-},v_{2}^{-},e^{-})\,.
\end{array}
\]

We observe that the sign of the branch coincides with the sign of the horizon curvature $k\neq 0$, so we can
simplify the notation by replacing $\pm\rightarrow k$.
In addition, $\Delta >0$ is equivalent to $\Lambda <\frac{1}{4Q^{2}}$ (which allows
both positive and negative $\Lambda $) and $\Delta >1$ is equivalent to $%
\Lambda <0$ (in agreement with the fact that $k=-1$ exists only for negative
$\Lambda $), so we conclude:

\begin{itemize}
\item[(\textit{i})]
When $\Lambda <0$, there are two
solutions corresponding to $k=\pm 1$,  given by Eqs.(\ref{solution u=0}), such that
the extremal black hole parameters are  $(v_1^k,v_2^k,e^k)$.

\item[(\textit{ii})]
When $0<\Lambda <\frac{1}{4Q^2}$,  there is only one
black hole solution with $k=1$ with the sign `$+$',
and
the extremal black hole parameters, given by Eqs.(\ref{solution u=0}), are $(v_1^+,v_2^+,e^+)$;

\item[(\textit{iii})] When $\Lambda =1/4Q^2$, there is no finite solution
for $v_1$.

\item[(\textit{iv})] The case $\Lambda =0$ can be reproduced from the limit $%
\Lambda \rightarrow 0$ of the positive branch of the solution (\ref{solution
u=0}).
\end{itemize}

To evaluate the entropy, it is useful to notice that, on the solution (\ref{solution u=0}),
the only dependence on the horizon geometry is through the
auxiliary function $f=\frac{kv_1}{2}$. Then extremum of the entropy
function, $ \mathcal{E}=2\pi\text{Vol}(\Sigma _k)(eQ-f)$, for the fixed charge and the values (\ref{solution u=0})
of the black hole parameters, is given by
\begin{equation}
S_{k}(Q)=\frac{2\pi Q^{2}\,\text{Vol}(\Sigma _{k})}{\sqrt{\Delta }+k}\,.
\end{equation}

For the spherical horizons ($k=1$), the horizon volume is
Vol$(\Sigma_1)=4\pi $ and the entropy becomes
\begin{equation}
S_+(Q)=\frac{8\pi ^{2}Q^{2}}{1+\sqrt{1-4\Lambda Q^{2}}}\,.  \label{Spm}
\end{equation}
Note that $S_{+}(Q)>0$ is always fulfilled.

When $\Lambda =0$, then the positive branch reproduces the well-known result (asymptotically flat, electric extremal RN black hole)
\begin{equation}
\left. S_{+}(Q)\right\vert _{\Lambda =0}=4\pi ^{2}Q^{2}=\frac{q^{2}}{4}\,.
\label{S0}
\end{equation}

For the negative branch ($k=-1$) and
non-compact horizons, the volume Vol$(\Sigma _{-1})$ is infinite, and we
look at the entropy density
\begin{equation}
s_{-}(Q)=\frac{S_{-}(Q)}{\text{Vol}(\Sigma _{-1})}=\frac{2\pi Q^{2}}{\sqrt{1-4\Lambda Q^{2}}-1}\,.
\end{equation}
It can be checked that $s_{-}(Q)>0$ is also satisfied because of negative values of the cosmological constant.

Now we turn to the case of hairy black holes.

\section{Hairy extremal black hole and the critical point}

For the study of phase transitions, the most interesting cases involve
non-linear Stueckelberg interaction $\mathcal{F}$ of the form (\ref{Stuckelberg}). The last equation in (\ref{eom}) implies that, for the
minimal coupling $a=0$, the only solution for the scalar field is $u=0$, for
which the black hole entropy was calculated in the previous section. When $a\neq 0$, then there are three solutions for the scalar parameter,
\begin{equation}
u=0\,,\qquad \,u=\pm \sqrt{\frac{2}{a}\left(\frac{v_1\,m^2}{e^2}-1\right) }\,.  \label{u}
\end{equation}

We found that there is more than one well-defined solution to the attractor equations stabilising the scalar field at the black hole horizon; this hints to the existence of \textit{basin of attractions} in the scalar manifold, possibly endowed with a non-trivial topology. Most of the results found in the literature are in the context of ungauged supergravity, without cosmological constant and without scalar potential; they have been usually associated to non-homogeneous scalar manifolds (see \textit{e.g.} \cite{BA11, BA12, BA13, BA21, BA22}), even if recent discoveries of multiple attractors \cite{Mandal1, Mandal2} seem to hold more in general. In our case, however, the basins of attraction appear in the spacetimes with non-vanishing $\Lambda$ and with a scalar potential. Choosing $\mathcal{F}$ suitably might lead to even more basins, increase the number of solutions and complicate their structure.

The existence of the basins of attraction would pertain to a possible phase transitions between different stable points in the near-horizon dynamics of the scalar field. To check this in our framework, we have to study the entropy near each of these points.

The case $u=0$ was analyzed in Section \ref{NonHairy}. When $u\neq 0$, the
equations of motion (\ref{u}) are invariant under the replacement $u\rightarrow -u$, so we can chose $u>0$.

Now we address the following question: Is there a critical point for which
two solutions $u=0$ and $u\geq 0$ co-exist? If the answer is yes, this would
be a possible branch point where a phase transition from one configuration to another might occur.

To answer this question, let us analyze the critical point limit
$u\rightarrow 0$ of the non-trivial solution in (\ref{u}). The critical point
exists only for the massive fields ($m\neq 0$). Then Eq.(\ref{u}) implies
\begin{equation}
v_{1c}=\frac{e_{c}^{2}}{m^{2}}\,.  \label{cr}
\end{equation}

We have to distinguish different horizon geometries. When $k=0$, then the
above constraint gives a consistent solution (\ref{planar}) only for the
particular scalar mass $m=\frac{1}{2}$, while the charge $Q$ remains
arbitrary. Thus, the co-existence line $u=0$ is not characterised by a
particular value of the charge (its critical value $Q_c$ does not exist), and an effective
potential of the system would not change drastically in this point with a change of $Q$,
implying that variations in the external parameter $Q$ would not be able to induce an instability of the system and trigger a phase transition.
Even though the scalar hair, in principle, still could develop in planar
black holes, they are not interesting to study in our context. Henceforth, we will consider the black holes with curved horizons only.

When $k \neq 0$, first two equations of (\ref{eom}) give that the critical point exists only
for $\Lambda \neq 0$ and $m^{2}\neq 1,\frac{1}{2}$, leading to
\begin{eqnarray}
v_{1c} &=&\frac{m^2-1}{\Lambda }\,,  \notag \\
e_{c} &=&\sqrt{\frac{m^2(m^2-1)}{\Lambda }}\,,  \notag \\
v_{2c} &=&\frac{k(m^2-1)}{\Lambda(2m^2-1)}\,.
\label{Critical fields}
\end{eqnarray}
The result does not depend on the strength of the scalar coupling $a\neq 0$.
Non-vanishing and positive $v_{ic}$ and $e_{c}$ exist only if
\begin{equation}
m^2>0\,,\qquad k(2m^2-1)>0\,,\qquad \frac{m^2-1}{\Lambda }>0\,.  \label{existence Cr}
\end{equation}
Note that the above conditions can be satisfied for both
positive and negative values of the cosmological constant. In addition, in
case of negative cosmological constant, the Breitenlohner-Freedman bound
that ensures the stability of scalar field in AdS$_{4}$ imposes $m^{2}>\frac{%
3\Lambda }{4}$ to the scalar mass, what is weaker than the inequalities (\ref%
{existence Cr}), so it is always satisfied.

The last unsolved equation (third in Eq.(\ref{eom})) shows that the critical
values of the parameters can be reached only for the critical charge
\begin{equation}
Q_{c}=\frac{v_{2c}e_{c}}{v_{1c}}=\frac{k}{2m^{2}-1}\sqrt{%
\frac{m^{2}\left( m^{2}-1\right) }{\Lambda }}\,.  \label{Qc}
\end{equation}

Compared to the original two $u=0$ solutions given by Eqs.(\ref{solution u=0}) where $Q$ is replaced by $Q_{c}$ we find that,
for both positive and negative branches, we reproduce the known critical results of the parameters,
\begin{equation}
v_1^{k}(Q_{c})=v_{1c}\,,\qquad v_2^{k}(Q_{c})=v_{2c}\,,\qquad e^{k}(Q_{c})=e_{c}\,.
\end{equation}
The critical entropy reads
\begin{equation}
S_c=\pi\text{Vol}(\Sigma _{k})\,\frac{m^2-1}{\Lambda k(2m^2-1)}>0\,,  \label{Sc}
\end{equation}
and it is a continuous function at $Q_c$ because
\begin{equation}
S_k(Q_c)=S_c\,.
\end{equation}

When $\Lambda =0$, then $m^{2}=1$ (and vice versa), and the solution is $v_{1c}=v_{2c}=Q^2$, $e_{c}=Q$ with the charge $Q$ which remains arbitrary.
Only $k=1$ is allowed. Non-existence of the critical charge is similar to the case $m^{2}=\frac{1}{2}$ of planar black holes discussed before and, as argued earlier, we will not look at these cases.

To summarize, we will discuss only the spacetimes with $\Lambda \neq 0$ and $k \neq 0$, when there exists a $Q_c$ as an isolated point, and analyze a behavior of the extremal hairy black hole in its vicinity. We will also explore whether a variation of electric charge could produce a phase
transition at this point using the second thermodynamic law, $\delta S\geq 0$.

\section{Black hole in the vicinity of the critical point}

A better physical understanding of the critical point of the extremal
black hole can be achieved by studying its behavior in the vicinity of the possible phase transition. We
focus on spherical and hyperbolic horizons, $k=\pm 1$, with $\Lambda \neq 0$, where the entropy is a continuous function of the electric charge at the critical point. As already mentioned, other cases do not lead to stable hairy black holes with isolated critical points.

The critical point separates $u=0$ and $u\neq 0$ extremal black
hole solutions given by Eqs.(\ref{Critical fields}) and (\ref{Qc}).
A near-critical behavior of the black hole is captured by a small parameter
\begin{equation}
\epsilon =Q-Q_{c}\,,
\end{equation}
where $\epsilon $ can be either positive or negative. We assume that $u$
vanishes when the charge approaches to the critical value so that,
near $Q_{c}$, the scalar field behaves as $u^{2}=A\,\epsilon^{\beta }+\cdots $,
where $\beta >0$ stands for a critical exponent.

In principle, different parameters $u^2$, $e$, $v_1$ and $v_2$ should have different critical exponents $\beta$, $\delta$, $\alpha$ and $\gamma$, respectively. However, as shown in Appendix \ref{Critical exponents}, the consistency of the equations of motion at the leading order, as well as
a requirement that the metric is a covariant function of $Q$ near the critical point, lead to the unique solution for the critical exponents $\beta=\delta=\alpha=\gamma=1$. Thus, all critical exponents are the same.

In particular, with this choice, the order parameter $u$ shows a typical mean-field behavior near the critical point, namely
\begin{equation}
u=\sqrt{A\,(Q-Q_{c})}+\mathcal{O}((Q-Q_{c})^{2})\,,
\end{equation}
which is universal, because it does not depend on the details of the system (the scalar mass, coupling constant and cosmological constant).

With this result at hand, we can solve the coefficients in the expansion of the parameters $u$, $v_{1}$,  $v_{2}$, and $e$,
using the method of successive approximations.
Let us assume that, for small $\epsilon $, the parameters behave as
\begin{eqnarray}
u &=&\sqrt{A\,\epsilon}+\sqrt{(\tilde{A}\,\epsilon)^3}+\cdots \,,  \notag \\
e &=&e_{c}+B\,\epsilon +\tilde{B}\,\epsilon ^{2}+\cdots \,,  \notag \\
v_{1} &=&v_{1c}+V\,\epsilon +\tilde{V}\,\epsilon ^{2}+\cdots \,,  \notag \\
v_{2} &=&v_{2c}+C\,\epsilon +\tilde{C}\,\epsilon ^{2}+\cdots \,. \label{near-critical}
\end{eqnarray}
The solution for the scalar field exists only if the quantities $A\,\epsilon$ and $\tilde{A}\,\epsilon$ are positive.

At linear order in $\epsilon $, above equations are solved in Appendix \ref{Critical exponents}, giving rise to the coefficients (\ref{B}), (\ref{A}), (\ref{C}) and (\ref{V}) which, for any $k\neq 0$ and $\Lambda \neq 0$, can be
written as
\begin{eqnarray}
A &=&-\sqrt{\frac{m^{2}-1}{\Lambda m^{2}}}\frac{4k\Lambda ^{2}(2m^{2}-1)^{2}%
}{\Lambda a-4(m^{2}-1)^{3}}\,,  \notag \\
B &=&\frac{k\Lambda a\,(2m^{2}-1)^{3}}{\Lambda a-4(m^{2}-1)^{3}}\,,  \notag
\\
C &=&\sqrt{\frac{m^{2}(m^{2}-1)}{\Lambda }}\,\frac{2\Lambda a+4(m^{2}-1)^{2}%
}{\Lambda a-4(m^{2}-1)^{3}}\,,  \label{ABCV} \\
V &=&\sqrt{\frac{m^{2}(m^{2}-1)}{\Lambda }}\,\frac{2k\Lambda
a\,(2m^{2}-1)^{2}}{\Lambda a-4(m^{2}-1)^{3}}\,.  \notag
\end{eqnarray}
The inequalities (\ref{existence Cr}) yield that the sign of $A$ depends only on the expression $k[4(m^{2}-1)^{3}-\Lambda a]$. For given $m^2$, $\Lambda$ and $a$, the sign of such a quantity is fixed. Thence, the condition $\mathrm{sgn}(A)(Q-Q_c)>0$ determines for which $Q$, above or below the critical point, a real $u$ exists and the black hole develops hair.

The entropy does not depend on the coefficients of second order. For
completeness, we write only the coefficient $\tilde{A}$ which has to be
positive,
\begin{equation}
\tilde{A}^{3}=\sqrt{\frac{k\Lambda ^{2}e_{c}}{\left[ 4(m^{2}-1)^3-\Lambda a\right]^3}}\frac{\alpha A}{8k\Lambda m(m^{2}-1) }\,,
\end{equation}
where $\alpha =32m^{2}(m^{2}-1)^{5}+4m^{2}\Lambda a(8m^{4}-2m^{2}-5)(m^{2}-1)^{2}+a^{2}\Lambda ^{2}(4m^{4}-3)$.

Replacing these values for the constants in the entropy function
(\ref{entropy f-on}), we find the entropy of the hairy black hole solution
near the critical point as
\begin{equation}
\left. S\right\vert _{u\neq 0}=S_c+8\pi ^2 e_{c}(Q-Q_c) +\mathcal{O}((Q-Q_c)^2)\,.  \label{Q>Qc}
\end{equation}

On the other hand, near the critical charge $Q_c$,
there are two extremal black hole solutions -- the hairy one whose entropy
is given by Eq.(\ref{Q>Qc}), and RN (A)dS with the entropy given by $S_{+}(Q)$ in Eq.(\ref{Spm}). When expanded
around the critical point, the RN solution becomes
\begin{eqnarray}
\left. S\right\vert _{u=0} &=&S_{+}(Q)=S_{c}+\left. \frac{dS_{+}}{dQ}%
\right\vert _{Q_{c}}\left( Q-Q_{c}\right) +\mathcal{O}((Q-Q_{c})^{2})  \notag
\\
&=&S_{c}+8\pi ^{2}e_{c}\left( Q-Q_{c}\right) +\mathcal{O}((Q-Q_{c})^{2})\,,  \label{Q<Qc}
\end{eqnarray}
what matches the hairy expression (\ref{Q>Qc}) up to quadratic terms.
Therefore, to distinguish two phases, the higher-power terms in the expansions (\ref{Q>Qc}) and (\ref{Q<Qc}) are needed.

After expanding the equations (\ref{eom}) at the $\epsilon ^{2}$ order, it is
straightforward to solve the algebraic equations and obtain the unique
solution in the coefficients $\tilde{A}$, $\tilde{B}$, $\tilde{C}
$ and $\tilde{V}$. We omit their writing in the text for the sake for
simplicity, especially because it can be shown that they do not contribute
to the $\epsilon ^{2}$ terms in $S(Q)$.

Plugging in this result into the entropy function, the entropy with $u\neq 0$ acquires the quadratic term.
By a comparison with the entropy with $u=0$, one obtains that
\begin{eqnarray}
\left. S\right\vert _{u\neq 0} &=&S_{c}+2\pi ^{2} \text{Vol}(\Sigma _{k}) e_{c}\epsilon
+k\pi \text{Vol}(\Sigma _{k})(2m^{2}-1)^{3}\,\omega \,\epsilon ^{2}+\mathcal{O}(\epsilon ^{3})\,,
\notag \\
\left. S\right\vert _{u=0} &=&S_{c}+2\pi ^{2} \text{Vol}(\Sigma _{k}) e_{c}\epsilon +
k\pi \text{Vol}(\Sigma _{k})(2m^{2}-1)^{3} \,\epsilon ^{2}+\mathcal{O}(\epsilon ^{3})\,.
\end{eqnarray}
We see that two values differ by the factor
\begin{equation}
\omega =\frac{\Lambda a}{\Lambda a-4(m^{2}-1)^{3}}\,.
\end{equation}
When $\omega \neq 1$ and if the RN (A)dS black hole develops hair above or below some critical charge, there is a discontinuity of the entropy at the critical point,
\begin{equation}
\lim_{Q\rightarrow Q_{c}^{+}}\frac{\partial ^{2}S}{\partial Q^2}\neq \lim_{Q\rightarrow Q_{c}^{-}}\frac{\partial ^{2}S}{\partial Q^2}\,.
\end{equation}
In order to establish which solution is more stable, we compare their
corresponding entropies,
\begin{equation}
\Delta S=\left. S\right\vert _{u\neq 0}-\left. S\right\vert _{u=0}=4\pi
^{2}(2m^{2}-1)^{3}\,(\omega -1) \,\epsilon ^{2}+\mathcal{O}(\epsilon ^{3})\,.
\end{equation}

According to the Second Law of Thermodynamics of black holes, a black hole
will develop hair if $\Delta S>0$, that is, $\omega >1$.
On the other hand, the hairy solution exists if $\mathrm{sgn}(A)(Q-Q_{c})>0$. These two
bounds give, respectively,
\begin{eqnarray}
\Lambda \left[ \Lambda a-4(m^{2}-1)^{3}\right]  &>&0\,,  \notag \\
k\left[ \Lambda a-4(m^{2}-1)^{3}\right] \left( Q-Q_{c}\right)  &<&0\,.
\label{bounds}
\end{eqnarray}
Therefore, given the cosmological constant and geometry of the horizon,
the inequalities (\ref{existence Cr}) fully determine an interval of allowed scalar
masses:

(\textit{a}) when $\Lambda >0$ and $k=1$, the masses lie in the interval $%
m^{2}>1$;

(\textit{b}) if $\Lambda <0$ and $k=1$, the masses lie in the interval $%
\frac{1}{2}<m^{2}<1$;

(\textit{c}) when $\Lambda <0$ and $k=-1$, the masses lie in the interval $%
0<m^{2}<\frac{1}{2}$. \medskip

Furthermore, the bounds (\ref{bounds}) determine for which values of the
coupling $a$ the phase transition would happen across the critical point.
Interestingly, in all three cases, the interaction has to be strong enough,
such that
\begin{equation}
a>\frac{4\left\vert m^{2}-1\right\vert ^{3}}{\left\vert \Lambda \right\vert }%
\,.  \label{a}
\end{equation}
Small or negative interactions do not favor the scalar hair.

In particular, in cases (\textit{a}) and (\textit{c}), the RN (A)dS black
hole exists above the critical point, for large charges. As the charge
decreases and passes through the critical point,  for $Q\leq Q_{c}$, the hairy
solution appears, which has larger entropy.

In the case (\textit{b}), the opposite situation happens. The RN AdS black
hole is favored for small charges and, as the charge
increases, the hair grows when the charge crosses the critical point, $Q\geq
Q_{c}$, also increasing the entropy of the configuration.

In all cases, there are phase transitions for either sign of $\Lambda $ and
any geometry of the horizon, provided the scalar coupling is strong enough.
We illustrate this on the examples.

As the first example corresponding to the case (\textit{a}), we take $\Lambda =3$, $k=1$, $m^{2}=2$,  $a=2$ and $\ell =1$, when the
critical parameters become
\begin{equation}
Q_{c}=\frac{1}{3}\sqrt{\frac{2}{3}}\,,\quad e_{c}=\sqrt{\frac{2}{3}}\,,\quad
v_{1c}=\frac{1}{3}\,,\quad v_{2c}=\frac{1}{9}\,.
\end{equation}
The scalar field exists for $Q\leq Q_{c}\,$,
\begin{equation}
u=3(54)^{1/4}\sqrt{Q_{c}-Q}+\frac{11043}{8}\,(24)^{1/4}\sqrt{(Q_{c}-Q)^{3}}+\cdots\,,
\end{equation}
the volume of the horizon is $\text{Vol}(\Sigma _1)=4\pi$, and the entropy close to the critical point has the form
\begin{equation*}
S(Q)=\left\{
\begin{array}{lll}
\frac{4\pi ^{2}}{9}+8\pi ^{2}\sqrt{\frac{2}{3}}\left( Q-Q_{c}\right) +324\pi
^{2}\,\left( Q-Q_{c}\right) ^{2}+\cdots \,,\quad  & Q\leq Q_{c} & \text{%
(hairy dS BH)\, ,} \\
\frac{4\pi ^{2}}{9}+8\pi ^{2}\sqrt{\frac{2}{3}}\left( Q-Q_{c}\right) +108\pi
^{2}\,\left( Q-Q_{c}\right) ^{2}+\cdots \,, & Q\geq Q_{c} & \text{(RN dS
BH)\, .}
\end{array}
\right.
\end{equation*}
As it can be seen, when the scalar exists, the hairy black hole has larger
antropy and the response function $S^{\prime \prime }(Q)$ is discontinuous
at $Q_{c}$.

As an example of the case (\textit{b}), we choose $\Lambda
=-3$, $k=1$, $m^{2}=3/4$, $a=1/12$ and $\ell =1$. The critical values of the
parameters in this case are
\begin{equation}
Q_{c}=\frac{1}{2}\,,\quad e_{c}=\frac{1}{4}\,,\quad v_{1c}=\frac{1}{12}
\,,\quad v_{2c}=\frac{1}{6}\,.
\end{equation}
Near the critical point, the scalar field behaves as
\begin{equation}
u=4\sqrt{Q-Q_{c}}+\frac{4}{3}\sqrt{(Q-Q_{c})^{3}}+\mathcal{\cdots \,}.
\end{equation}
Thus, when $Q\geq Q_{c}$, there are two solutions ($u=0$ and $u\neq 0$),
whereas for $Q\leq Q_{c}$, there is only RN AdS solution ($u=0$). The
near-critical entropy reads
\begin{equation*}
S(Q)=\left\{
\begin{array}{lll}
\frac{2\pi ^{2}}{3}+2\pi ^{2}\left( Q-Q_{c}\right) +\frac{2\pi ^{2}}{3}
\,\left( Q-Q_{c}\right) ^{2}+\cdots \,,\quad  & Q\geq Q_{c} & \text{(hairy AdS BH)\thinspace ,} \\
\frac{2\pi ^{2}}{3}+2\pi ^{2}\,\left( Q-Q_{c}\right) +\frac{\pi ^{2}}{2}
\,\left( Q-Q_{c}\right) ^{2}+\cdots \,, & Q\leq Q_{c} & \text{(RN AdS BH)\thinspace ,}
\end{array}
\right.
\end{equation*}
showing a discontinuity of $S^{\prime \prime }$ at $Q_{c}$.

The last example corresponds to the hyperbolic horizon case (\textit{c}).
Choosing $\Lambda =-3$, $k=-1$, $m^{2}=1/4$, $a=1$ and $\ell =1$, we obtain
the critical parameters%
\begin{equation}
Q_{c}=\frac{1}{2}\,,\quad e_{c}=\frac{1}{4}\,,\quad v_{1c}=\frac{1}{4}
\,,\quad v_{2c}=\frac{1}{2}\,,
\end{equation}
and the scalar field close to the critical point exists only if $Q\leq Q_{c}$,
\begin{equation}
u=4\sqrt{\frac{3}{7}\,(Q_{c}-Q)}-\frac{148}{7}\sqrt{\frac{3}{7}%
\,(Q_{c}-Q)^{3}}+\mathcal{\cdots \,}.
\end{equation}
The entropy density close to $Q_c$ behaves as
\begin{equation*}
s(Q)=\left\{
\begin{array}{lll}
\frac{\pi }{2}+\frac{\pi }{2}\,\left( Q-Q_{c}\right) +\frac{2\pi }{7}%
\,\left( Q-Q_{c}\right) ^{2}+\cdots \,,\quad  & Q\leq Q_{c} & \text{(hairy
AdS BH)\thinspace ,} \\
\frac{\pi }{2}+\frac{\pi }{2}\,\left( Q-Q_{c}\right) +\frac{\pi }{8}\,\left(
Q-Q_{c}\right) ^{2}+\cdots \,, & Q\geq Q_{c} & \text{(RN AdS BH)\thinspace .}%
\end{array}
\right.
\end{equation*}
It is clear that the extremal black hole, under the aforementioned
conditions, undergoes a phase transition at $Q_{c}$. To the best of our
knowledge, this is the first observation of such a phenomenon for the class
of black holes under consideration.

\section{Conclusions}

We studied phase transitions of extremal black holes in (A)dS$_4$ gravity
that arise due to variations of electric charge.
The role of order parameter is played by a complex Stueckelberg scalar field.
We showed that the (necessarily massive) scalar field can couple to a RN black hole at zero temperature
only if the space-time has a non-vanishing cosmological constant, and if a non-minimal coupling to gravity is present; moreover, the RN black hole should carry no magnetic charge at all. If the cosmological constant is negative, only spherical and hyperbolic geometries of the horizon are admitted, and there are no black branes. We were able to analytically show that this system possesses a critical point characterized by the critical electric charge.

The critical charge depends only on the scalar mass and on the cosmological constant. For spherical  and hyperbolic horizons, on one side of the critical point, there is only one possible solution, the extremal RN (A)dS black hole. On the other side of the critical point, there are two possible solutions: RN (A)dS, and a hairy RN (A)dS black hole which has larger entropy. Thus, the black hole is more stable if it develops scalar hair, implying that a change in the electric charge around the critical point would produce a zero temperature phase transition. This is characterized by a continuous $S$ and $\partial S / \partial Q$  at the critical point, but discontinuous second derivative of the entropy with respect to the electric charge, $\partial^2 S / \partial Q^2$. These results were obtained using the entropy function formalism, so they apply for any bulk hairy black hole solution.

For planar horizons, there exists a (non-hairy) RN AdS black hole solution, but there is no a critical point where the hair should start developing, which could be reached by a change of the external parameter, $Q$, so standard phase transitions in this case were not found. A similar situation occurs when $\Lambda =0$. The phase transition was observed, therefore, only for $\Lambda \neq 0$  and  $k\neq 0$.

The co-existence of different well-defined solutions to the attractor
equations stabilising the scalar field at the black hole horizon hints for
the possible relevance of \textit{basins of attractions} \cite{BA11,BA12,BA13} in the
framework under consideration. Might the phase transitions be driven by the
``area codes'' \cite{BA21,BA22} selecting the different basins of
attraction in the near-horizon dynamics of the scalar field? Exploiting
AdS/CFT correspondence, what is their interpretation and relevance in the
dual CFT? We hope to report on attempts to answer these questions in future
works.

Moreover, we leave for further future investigation the analysis of the
possibility to regard the action (\ref{action}) as the purely bosonic sector of a
certain $\mathcal{N}=1$, $D=4$ supergravity theory coupled to one chiral
multiplet and one vector multiplet, with suitable non-Abelian gauging
consistent with the Stueckelberg coupling (the corresponding holomorphic
Killing vector being given by $ \partial_{\varphi}$) and a related scalar potential.
In the model (\ref{action}) under consideration, there is only vector field, and its
kinetic vector term is of Maxwell type, corresponding to the limit in which
the holomorphic kinetic coupling is a real constant (\textit{minimal coupling}). Here, we limit ourselves to remark that the minimal supersymmetric
extension of chaotic inflation \cite{infl-1, infl-2} is described by the
supersymmetric Stueckelberg model coupled to $\mathcal{N}=1$ supergravity
(corresponding to a flat K\"{a}hler space); this fact was first pointed out
in \cite{infl31,infl32} and proved to be the zero-curvature limit \cite{infl41,infl42} of
a continuous class of $SU(1,1)/U(1)$ gauged sigma models \cite{infl31,infl32} named
\textit{alpha attractors} \cite{infl41,infl42}.

Finally, it would be interesting to consider further generalizations,
introducing more vector \textit{and/or} scalar fields, with a non-trivial
(possibly holomorphic) kinetic vector matrix, and subsequently trying to
embed them into $\mathcal{N}\geqslant 2$-extended supergravity theories\footnote{The systematics of reduction from $\mathcal{N}>1$ to $\mathcal{N}=1$
theories with \textit{minimal coupling} has been investigated in \cite{Deg-Type-E7}, and further necessary conditions have been obtained in \cite{ADF-1, ADF-2}.}.

\acknowledgments

The authors would like to thank (in alphabetic order) Marco Astorino, Patrick Concha, Alessandra Gnecchi and Ricardo Troncoso for
enlightening discussions and useful correspondence.
This work was partially supported by the Chilean FONDECYT Project No.1170765
and the VRIEA-PUCV Grants No.039.428/2017 and No.123.752/2017.
P.Q.L. is a PUCV scholarship holder.

\appendix

\section{Critical exponents}

\label{Critical exponents}

Here we provide the proof that the parameters $u^{2}$, $e$, $v_{1}$ and $v_{2}$ have the
same critical exponents.

We seek for a solution of the equations of motion (\ref{eom}) for $k = \pm 1$ in
the form of the power-law in the small quantity $\epsilon =Q-Q_{c}$, that is,
\begin{eqnarray}
u^{2} &=&A\,\epsilon ^{\beta }+\cdots \,,  \notag \\
e &=&e_{c}+B\,\epsilon ^{\delta }+\cdots \,,  \notag \\
v_{1} &=&v_{1c}+V\,\epsilon ^{\alpha }+\cdots \,,  \notag \\
v_{2} &=&v_{2c}+C\,\epsilon ^{\gamma }+\cdots \,.
\end{eqnarray}
Positive numbers $\alpha $, $\beta $, $\gamma $ and $\delta $ denote the
critical exponents, and $A$, $B$, $V$, $C$ are non-vanishing coefficients. It is useful to
eliminate the mass parameter from the field equations by means of the identity $m^{2}=\frac{e_{c}^{2}}{v_{1c}}$.
The leading order of the equations (\ref{eom}) thus becomes
\begin{eqnarray}
0 &=&\dfrac{2e_{c}v_{2c}}{v_{1c}^{2}}\,B\,\epsilon ^{\delta }
+\left( \dfrac{e_{c}^{2}}{v_{1c}^{2}}+\Lambda \right) C\epsilon ^{\gamma }
-\frac{2e_{c}^{2}v_{2c}V}{v_{1c}^{3}}\,\epsilon ^{\alpha }
+\frac{e_{c}^{2}v_{2c}}{v_{1c}}\,A\,\epsilon ^{\beta }\,,  \label{1} \\
0 &=&\frac{2e_{c}}{v_{1c}}\,B\,\epsilon ^{\delta }-\left( \Lambda
 +\frac{e_{c}^{2}}{v_{1c}^{2}}\right) V\,\epsilon ^{\alpha }\,,  \label{2} \\
\epsilon  &=&\frac{e_{c}}{v_{1c}}\,C\,\epsilon ^{\gamma }
+\dfrac{v_{2c}}{v_{1c}}\,B\,\epsilon ^{\delta }-\frac{v_{2c}e_{c}}{v_{1c}^{2}}\,V\,\epsilon
^{\alpha }+v_{2c}e_{c}\,A\,\epsilon ^{\beta }\,,  \label{3} \\
0 &=&\frac{2e_{c}^{2}}{v_{1c}}\,V\,\epsilon ^{\alpha }-4e_{c}B\,\epsilon
^{\delta }-ae_{c}^{2}\,A\,\epsilon ^{\beta }\,.  \label{4}
\end{eqnarray}
The finite order of the equations was cancelled out because it corresponds to the critical point. In that way, explicit dependence on $k$ drops out from Eq.(\ref{1}), and it enters implicitly only through the critical parameter $v_{2c}$, as seen from Eq.(\ref{Critical fields})). Also, the coupling $a$ appears only in the last term of the last equation.

From Eq.(\ref{2}) it is clear that $B$ and $V$ are non-vanishing only if $\delta =\alpha $, in which case
\begin{equation}
B=\frac{v_{1c}}{2e_{c}}\left( \Lambda +\frac{e_{c}^{2}}{v_{1c}^{2}}\right)
\,V\,.  \label{B}
\end{equation}%
Replacing these results in Eq.(\ref{4}) leads to
\begin{equation}
0=2v_{1c}\Lambda \,V\,\epsilon ^{\alpha }+ae_{c}^{2}\,A\,\epsilon ^{\beta
}\,.
\end{equation}
Again, the coefficients are non-vanishing only if $\alpha =\beta $,
giving rise to
\begin{equation}
A=-\frac{2v_{1c}\Lambda }{ae_{c}^{2}}\,V\,.  \label{A}
\end{equation}%
With these results, Eq.(\ref{1}) becomes
\begin{equation}
0=\dfrac{v_{2c}}{v_{1c}}\left( \Lambda -\frac{e_{c}^{2}}{v_{1c}^{2}}-\frac{%
2v_{1c}\Lambda }{a}\right) \,V\,\epsilon ^{\alpha }+\left( \dfrac{e_{c}^{2}}{%
v_{1c}^{2}}+\Lambda \right) C\epsilon ^{\gamma }\,.
\end{equation}%
Since the coefficients in the brackets are not zero (as can be checked out directly by
using the expressions (\ref{Critical fields})), the above equation is
consistent only if $\gamma =\alpha $. Then we solve the coefficient
\begin{equation}
C=\dfrac{v_{2c}}{v_{1c}}\frac{\frac{e_{c}^{2}}{v_{1c}^{2}}+\frac{%
2v_{1c}\Lambda }{a}-\Lambda }{\dfrac{e_{c}^{2}}{v_{1c}^{2}}+\Lambda }\,V\,.
\label{C}
\end{equation}
Using all known expressions, the last equation (\ref{3}) acquires the form
\begin{equation}
\epsilon =\frac{v_{2c}}{2e_{c}}\left( \Lambda -\frac{e_{c}^{2}}{v_{1c}^{2}}-%
\frac{4v_{1c}\Lambda }{a}\right) \,V\epsilon ^{\alpha }+\dfrac{e_{c}v_{2c}}{%
v_{1c}^{2}}\frac{\frac{e_{c}^{2}}{v_{1c}^{2}}+\frac{2v_{1c}\Lambda }{a}%
-\Lambda }{\dfrac{e_{c}^{2}}{v_{1c}^{2}}+\Lambda }\,V\epsilon ^{\gamma }\,.
\label{last EQ}
\end{equation}

So far, we have that three critical exponents are equal, $\delta =\alpha =\beta $. It remains to determine $\gamma$.
 To have $V\neq 0$, it has to hold
either $\gamma =\alpha =1$, or $\gamma >\alpha =1$, or $1=\gamma <\alpha $.
All three cases are mathematically allowed, but only the case $\gamma =\alpha$ is physically sensible,
as these critical exponents define a behavior of the metric
(i.e. $v_{1}$ and $v_{2}$) near the critical point, and one would expect in
gravity that\ the metric remains covariant at any $Q$, and therefore it
tends to the critical point in a\ covariant way, $g_{\mu \nu }=g_{c\mu \nu
}+H_{\mu \nu }\,\epsilon ^{\alpha }$.

As a consequence, only the first case is physically allowed, and all
critical exponents are equal to one ($\delta =\alpha =\beta =\gamma =1$).
Then we find%
\begin{equation}
V=\frac{\frac{2e_{c}}{v_{2c}}\left( \dfrac{e_{c}^{2}}{v_{1c}^{2}}+\Lambda
\right) }{\left( \Lambda -\frac{e_{c}^{2}}{v_{1c}^{2}}-\frac{4v_{1c}\Lambda
}{a}\right) \left( \dfrac{e_{c}^{2}}{v_{1c}^{2}}+\Lambda \right) +\dfrac{%
2e_{c}^{2}}{v_{1c}^{2}}\left( \frac{e_{c}^{2}}{v_{1c}^{2}}+\frac{%
2v_{1c}\Lambda }{a}-\Lambda \right) }\label{V}\,,
\end{equation}%
from where other constants (\ref{B}), (\ref{A}) and (\ref{C}) can be
expressed in terms of the critical parameters. Replacing the critical values
from (\ref{Critical fields}), one finds the form (\ref{ABCV}) presented in
the main text, where all critical coefficients are equal, for both spherical and hyperbolic horizons.

\end{document}